
\magnification=1200
\vsize=22truecm
\hsize=15truecm \tolerance 1000
\parindent=0pt
\baselineskip = 15pt
\lineskip = 1.5pt
\lineskiplimit = 3pt
\font\mybb=msbm10
\def\bb#1{\hbox{\mybb#1}}
\voffset=0pt
\def\la{\langle}
\def\ra{\rangle}
\parskip = 1.5ex plus .5ex minus .1ex
{\nopagenumbers
\rightline{KCL-TH-93-2}
\vglue 1truein
\centerline{THE COVARIANT SCATTERING AND COHOMOLOGY}
\centerline{OF $W_3$ STRINGS}

\bigskip
\centerline{Michael Freeman}
\centerline{\&}
\centerline{Peter West}
\bigskip
\centerline{Department of Mathematics}
\centerline{King's College}
\centerline{Strand}
\centerline{London WC2R 2LS}
\bigskip
\centerline{February 1993}
\vskip 1truein
\centerline{Abstract}
\medskip
A general formalism for covariant $W_3$ string scattering is given.
It is found necessary to use screening charges that are constructed
from the $W_3$ fields including ghosts.  The scattering amplitudes
so constructed contain within them Ising model correlation functions
and agree with those found previously by the authors.
Using the screening charge and a picture changing operator, an
infinite number of states in the cohomology of Q are generated
from only three states.  We conjecture that, apart from discrete
states, these are all the states in the cohomology of Q.
\vfil
\eject}

{\bf Introduction}

The existence of $W$-algebra [1,2] extensions of the Virasoro
algebra has led to a number of new developments.  One of these has
been the discovery of new string theories based on these extensions
of the (super) Virasoro algebra.  An important early result was
the construction of the BRST charge
for the $W_3$ algebra, which was found to be nilpotent for a matter
central charge of $c=100$ and to imply intercept 4 for standard
ghost number states [3].  This work was used to discuss the
possible properties of $W_3$ string theories in references [4]
and [5].  Unlike Virasoro representations, two $W_N$ algebra
representations cannot be added to obtain a third representation.
The most commonly used representations of the $W_N$ algebras are
those of reference [6], which involve $N-1$ free scalar fields
and can have any value of the central charge.  A variant of these
representations was found [7] which replaced one of these scalar
fields by any number of scalar fields $x^\mu,\,\mu=0,1,\dots ,D-1$.

These representations with $c=100$ were then used to construct
$W_3$ strings [8,9,10].  The physical states of the bosonic
string were found long ago[11], but we now realize that they
can be thought of as
being the cohomology of the BRST operator $Q$, subject to a ghost
number constraint.  We recall that the cohomology of $Q$ [12]
consists of the states $\vert\psi^x\ra\vert\downarrow \ra$ and
$\vert\psi^x\ra \,c_0\vert\downarrow \ra$,
where $\vert\psi^x\ra$ depends only on
$x^\mu$ and $\vert\downarrow\ra=c_1\vert 0\ra$, with $\vert 0\ra$ being
the $SL(2,\bb{R})$ invariant vacumm.  There are also two further states
with zero momentum.

For a $W_3$ string we also regard the physical states as being
given by the cohomology of $Q$, subject to a suitable ghost number
constraint.  It was pointed out in references [4,5] that if
the physical states had a ghost vacuum of standard type, that is
they are of
the form $\vert\psi^{x,\varphi}\ra\vert\downarrow\ra$ where
$\vert\downarrow \ra=c_1e_1e_2\vert 0\ra$, then the $x^\mu$-$\varphi$
dependent part $\vert\psi^{x,\varphi}\ra$ would be subject to the
constraints
$$
\eqalign{(L^m_0-4)\vert\psi^{x,\varphi} \ra=&0,\quad
W^m_0\vert\psi^{x,\varphi} \ra=0,\cr
L^m_n\vert\psi^{x,\varphi} \ra=&0,\quad
W^m_n\vert\psi^{x,\varphi}\ra=0,\quad n\geq 1.\cr}\eqno(1.1)
$$
We give  details of our conventions in section 2.

Included amongst the states satisfying the conditions
of equation (1.1) are physical states having the form [8,9,10]
$$
\vert\psi^{x,\varphi}\ra = \vert\psi^x\ra
\vert 0,\beta\ra\vert\downarrow\ra,\eqno(1.2)
$$
where $\vert 0,\beta \ra$, is a state with $\varphi$ momentum
equal to $\beta$ and
no $\varphi$ oscillators.  Such states will satisfy the conditions (1.1)
provided that the state $\vert\psi^x\ra$, which
depends on $x^\mu$ alone, satisfies the conditions
$$
L^x_n\vert\psi^x\ra=0,\ n\geq 1,\quad(L^x_0-a)\vert\psi^x\ra=0,\eqno(1.3)
$$
where $a=1$ for $\beta={8iQ/7}$ and ${6iQ/7}$, and
$a={15/16}$ for $\beta=iQ$.  The above values of $\beta$ are
in fact the only ones allowed by the on-shell conditions of equation
(1.1).  We will refer to such states as intercept 1 or
${15/16}$ states.  A systematic study of the physical states of
equation (1.1) at levels up to and including 2 was undertaken in
reference [13].  It was found that any state that had the form
given in equation (1.1) and that contained
$\varphi$ oscillators was null, and, by examining all other
null states, the count of physical degrees of freedom at these
levels was found.  It thus became clear that the non-null physical
states were of the form of equation (1.2) and that the open $W_3$
string had only one massless particle, a photon.

Already it had been noticed for the $W_N$ string, as a matter of
phenomenological number matching, that the allowed values
of the intercepts $a$ referred to above were
related to the weights of some of the primary fields of the
minimal conformal models [8,9], and also that the fields and ghosts of the
$W_N$ string, with the exception of $x^\mu$ and the $b,c$ ghosts, had
the central charge of these models [8]. For $W_3$ this
observation amounts to the fact that $0=1-1$ and ${1/16}=1-{15/16}$ were
weights of Ising primary fields, and that the $\varphi, d, e$ system has a
central charge $c={1/2}$.

For the bosonic string it is only necessary to consider
physical states built on the
standard ghost vacuum $\vert\downarrow \ra$, since, in addition to
these states, the cohomology of
$Q$ consists of only a copy of these states built on
$c_0\vert\downarrow \ra$ and two so-called discrete states having
fixed momenta. For the $W_3$ string, however,  it emerged in two
ways that the situation was not so simple.
Physical states having ghost number 2
were first found in reference [14] in the context of the two
scalar $W_3$ string.  These authors realized that such states were,
in the sense of number matching discussed above, associated with
Ising weight ${1/2}$ states.  It is known [15] that the
two-dimensional bosonic strings have discrete physical states, that
is states with fixed momenta, occurring at a variety of different
ghost numbers.  The analogue of this phenomenon was discussed for
the $W_3$ string in reference [16], and these authors also
found two examples of ghost number 1, level one physical states
having continuous momenta in the 3 scalar $W_3$ string.

In a separate development [17] it was found that scattering of
$W_3$ strings was not consistent with states corresponding to
intercepts 1 and ${15/16}$ alone, since factorization of the
amplitude for 4 of the ${15/16}$ intercept strings revealed an
infinite number of intermediate states associated with intercept
${1/2}$.  In reference [18] all the physical states of
the type of equation (1.3) were classified by constructing a
spectrum generating algebra.  This consisted of the operators
$B^i_n,1=1,\dots ,D - 1$ and $B_n$, where the latter obeyed a
Virasoro algebra with central charge ${1/2}$.  It was shown
[18] that the count of states was given by the Ising model
characters $\chi_h$, where $h=1-a$, and that the states
had positive norm
and so obeyed a no-ghost theorem.  These results allowed a discussion
of modular invariance.  It emerged [18] that the $W_3$ string
was not modular invariant for the intercept ${15/16}$ and 1
states alone, but required states with a count given by a
character $\chi_{1/2}$.  It was realised that states of the
form of equation (1.3) for $a={1/2}$ would have precisely this
count and that such states did indeed
belong [18] to the cohomology of
$Q$ at ghost number 1.  It also emerged that the cohomology of
$Q$ contained complete
copies of the intercept ${15/16}$ and ${1/2}$ states.

Taking the above facts into account, it was conjectured [18]
that the cohomology of $Q$ contained only intercept $1,{15/16}$
and ${1/2}$ states, together with copies of them and possible
discrete states.  We will refer to this conjecture as the spectrum
conjecture.  A review of $W_3$ string theory including its
ghosts and physical states is given in section 2.

In reference [17] a formula for the scattering of any $W_3$
strings was derived using the group theoretic approach.
This approach stemmed from a desire to formulate a method of
computing string amplitudes that utilized the minimum amount of
machinery and
assumptions.  It was this that enabled us to work with a reduced
subspace of the full Hilbert space using properties such as the
null state structure of the states and $W_3$ properties of the
vertices that followed from the full $W_3$ theory.  These results
[17], which showed that $W_3$ string scattering amplitudes
contain within them Ising model correlators, are reviewed in
section 3.  Further explicit evaluations of particular scattering
amplitudes are given in section 3, so that they may be compared with
other derivations.

It was suggested in reference [17] that it would be worthwhile
to recover these results from a covariant formalism.  This is the
central task of this paper.  The covariant formalism is taken to
mean that which is often embodied in the conformal field theory approach
to string scattering.  This  method [19] constructs the
amplitudes as vacuum expectation values, with respect to the
$SL(2,\bb{R})$-invariant vacuum, of
BRST-invariant vertex operators (such as $ce^{ip\cdot x}$ for
the bosonic string) and of BRST-invariant integrated vertices
obtained by acting with the operator $\int dz\oint_z dv b(v)$
on a BRST-invariant operator $V(z)$; such integrated vertices have
ghost number one less than the original vertex $V$.
In the case of the superstring it is also necessary to use
picture changing operators constructed
from the bosonized ghosts.  These building blocks must be
assembled in such a way that the final amplitude is BRST invariant.
The bosonic and superstring theory tree amplitudes have been known
since the earliest days of string theory, and the rules used
originally proved a very useful guide to constructing rules of the
covariant formalism.

For a new string theory such as a $W_3$ string the bosonic and
superstring rules provide a useful guide and we will need all the
above building blocks.  There is, however, an element of guesswork
in this, and as such it is important to verify that the final
scattering amplitudes are consistent as a check on a given set of
rules.  We find in section 4 that for the $W_3$ string one must
extend the usual covariant scattering rules to include screening
charges, that is, objects that are integrals of a vertex operator
that is not associated with an external string.  A departure of
some kind is perhaps to be expected in view of the role which will
be played by the bosonic $W$-moduli, once this phenomenon is
understood.

The screening charges, which are constructed from the fields of
the $W_3$ string, play a role analogous to the screening
operators used in the
Feigin-Fuchs [20] formulation of minimal conformal models.  The
scattering of 4 intercept ${15/16}$ strings is explicitly
evaluated in section 4 and found to be in agreement with the
result of reference [17], where it was found that it contained
the Ising model
correlation function for 4 weight ${1/16}$ primary fields.

In section 5 the cohomology of $Q$ is examined.  We show that,
starting from one BRST vertex operator $V(a,0)$ for each
intercept $a=1,{15/16},{1/2}$, we can act with the
screening operator $S$ and picture changing operator $P$ in
suitable ways to obtain whole families of BRST invariant vertex
operators $V({15/16},m)$, $V(1,m)$, $\bar V(1,m)$, $V({1/2},m)$,
$\bar V({1/2},m)$ for $m\in{\bb Z}$, together
with extra states given
by the action of $P$ on these.  For example
$V({15/16},m)$ is given by
$$
V({15/16},m)=(S^2P)^mV({15/16},0).
$$
We extend our spectrum conjecture in the light of this construction.

In section 6 we give a complete set of rules for covariant $W_3$
string scattering.  Any scattering amplitude can be built from the
three vertices $V(a,0)$, the screening charge $S$, the picture
charging operator $P$ and the operation of
$\int dz\oint_z dv\, b(z)$
on a vertex.  We explain how to assemble these building
blocks and give a number of examples.

In references [21] and [22] an attempt was made to give a
formulation of covariant $W_3$ scattering using only the usual
rules found in the bosonic string and superstring.  In this attempt many
scattering amplitudes vanished, including that for four intercept
${15/16}$ tachyonic strings.  The authors noted that this
disagreed with the results of reference [17] and concluded that
reference [17] was incorrect.  The factorization and duality
properties of the amplitude for four intercept ${15/16}$
tachyonic strings were discussed in detail in reference [17] and
the amplitude was shown to factorize into two three point couplings
consistent with the Ising model fusion rules.  In section 7 we
explain why the rules of references [21,22] lead to string amplitudes
violating unitarity and the assumptions of S-matrix theory, as well
as assumptions more specific to string theory such as duality

{\bf 2. The covariant $W_3$ string}

In this section we summarize those features of the $W_3$ string,
including its ghosts, that will be required later in the paper.

The covariant formulation of the $W_3$ string involves the scalar
fields $\varphi, x^\mu$ for  $\mu = 0,1,..., D-1$, reparameterization
ghosts $b$ and $c$, and $W_3$ transformation ghosts $d$ and $e$.  The latter
ghosts have spins $3$ and $-2$ respectively.  The corresponding
energy-momentum tensor $T^{tot}$ and $W_3$-current $W^{tot}$ are
of the form
$$
T^{tot} = T^m + T^{gh} , \quad W^{tot} = W^m + W^{gh} \eqno(2.1)
$$
where
$$
T^m  = T^\varphi + T^x \eqno(2.2)
$$
and
$$
\eqalignno{
T^\varphi & = -{1\over 2}{(\partial\varphi)}^2 -
  Q \partial^2\varphi,&(2.3)\cr
T^x & = -{1\over 2} \partial x^{\mu} \partial x_{\mu}
  - \alpha_{\mu}\partial^2 x^{\mu}, & (2.4)\cr
W^m & =- {2i\over \sqrt{261}}\left[{1 \over 3} { (\partial \varphi)}^3 +
  Q \partial\varphi \partial^2 \varphi
  +{1 \over 3} Q^2 \partial^3 \varphi +
  2 \partial\varphi T^x + Q \partial T^x\right], &(2.5)\cr
T^{gh} &  = - 2 b\,\partial c - \partial b\,c - 3d\, \partial e -
2 \partial d\, e , &(2.6)\cr
W^{gh} & = - \partial d\, c - 3 d\, \partial c -{8\over 9.27} [\partial
(b\,e\, T^m) + b\,\partial e\, T^m]\cr
& + {25\over 6.9.27} (2 e\, \partial^3 b + 9 \partial e\, \partial^2 b
  + 15 \partial^2 e\, \partial b + 10 \partial ^3 e\, b). &(2.7)
}
$$
Here the background charge $Q$ is given by $Q^2 = 49/8$, and $\alpha$ is
such that $T^x$ has central charge $51/2$.
The BRST charge $Q$ is given by [3]
$$
Q = \int dz \, j^{BRST}, \eqno (2.8)
$$
where
$$
j^{BRST} = c(T^m + {1\over2} T^{gh}) + e(W^m + {1\over 2} W^{gh}).\eqno (2.9)
$$
(There will be no confusion between the background charge $Q$ and the
BRST charge $Q$.)
Some useful relations are
$$
T^{tot}(z) = \left\{ Q, b(z)\right\} , \quad W^{tot} (z) =
\left\{Q, d(z) \right\}, \eqno (2.10)
$$
as a consequence of which
$$
[Q, T^{tot}(z)] =  \left [Q, W^{tot}(z)\right ] = 0. \eqno (2.11)
$$
It will be useful to discuss the various possible vacua associated
with the ghosts.  The natural vacuum, $\vert 0\ra$, of the ghost system is
that for which $e(z) = \sum_n e_{-n} z^{n+2}$ and
$d(z) = \sum \limits _n d_{-n} z^{n-3}$ are well defined at $z = 0$.
This requires [19]
$$
\eqalign{
e_n \vert 0\ra &= 0, \quad n\ge3,\quad  d_n \vert 0\ra = 0,\quad n\ge -2\cr
c_n \vert 0\ra &= 0, \quad n\ge 2,\quad b_n \vert 0\ra = 0, \quad n\ge -1.\cr}
\eqno(2.12)
$$
We can construct other vacua by acting on $\vert 0\ra$ with $e_n$
for $n=0,\pm 1, \pm 2$ and with $c_n$ for  $n=0, \pm 1$. One
of the most useful is
$$
c_1 e_1 e_2 \vert 0 \ra \equiv  \vert \downarrow \ra, \eqno (2.13)
$$
which is annihilated by $e_n, c_n$ for $n \ge 1$ and $b_n, d_n$ for
$n\ge 0$.  In
terms of the conformal fields we may express the relation between
the two states as
$$
c\,\partial e\, e \vert 0 \ra = \vert \downarrow \ra, \eqno (2.14)
$$
where $c\, \partial e\, e$ is understood to be evaluated at $z =0$.
Similar formulae hold for the other vacuum states.

In order to gain a non-zero vacuum expectation value with respect to the
state $\vert 0\ra$ we must have 3 factors of $c$ and 5 of $e$. We set
$$
\la 0\vert c_{-1}c_0 c_1 e_{-2} e_{-1} e_0 e_1 e_2 \vert 0\ra  =
{1\over 576} \la 0\vert \partial^2 c\, \partial c\, c
\partial^4 e\, \partial^3 e\,
\partial ^2 e\, \partial e\, e\vert 0\ra  = 1 \eqno (2.15)
$$
It is often useful to bosonize the ghosts.  We adopt the well known
rules [19]
$$
c = e^{i\sigma},\quad b = e ^{-i\sigma}, \quad e = e^{i \rho},
\quad
d = e^{-i \rho}, \eqno (2.16)
$$
where $\sigma (z) \sigma (w) = - \ln(z -w)$ and similarly for
$\rho$ in order to give the correct operator product relations for
the ghosts.  The fields $\sigma$ and $\rho$ have energy-momentum
tensors
$$ T^\sigma = - {1\over 2}( \partial \sigma )^2 + {3 i\over 2}
\partial^2 \sigma,
\quad T^\rho = -{1\over 2} (\partial\rho )^2 + {5 i \over 2}
\partial^2 \rho. \eqno (2.17)
$$
We recognise that these background charges are compatible with
equation (2.15) once we use formulae such as
$$ \partial c\, c = : \partial e ^{i\sigma} e ^{i\sigma}:(z)
= \oint_z dw
{\partial_w\left (e^{i\sigma(w)} e^{i\sigma}(z) \right)\over w-z} =
e^{2i\sigma} (z). \eqno(2.18)$$
We take the vacuum $\vert \downarrow \ra $ to have ghost number
0, $e$ and $c$ to have ghost number $+1$ and $d$ and $b$ to have ghost
number $-1$.

We now summarize some of the states known to be in the cohomology
of Q.  The first such states that were found [4,5,8,9] were those of
so-called standard ghost type, which were based on the
$\vert\downarrow \ra$ vacuum and were of the form
$$
\vert \psi \ra\vert \downarrow \ra, \eqno (2.19)
$$
where $\vert \psi \ra$ is made from the $\alpha_n$ and $\alpha^\mu_n$
oscillators of $\varphi$ and $x^\mu$ respectively acting on their
usual vacua.  These states are annihilated by $Q$ if
$$
(L^m_0 - 4 ) \vert \psi \ra = 0 = W^m_0 \vert\psi \ra, \quad L^m_n
\vert \psi \ra = W^m_n \vert \psi \ra = 0, \quad n\ge 1. \eqno (2.20)
$$
It was shown [13] that such physical states, up to level 2 at
least, were either null or of the form
$$
e^{i\beta \varphi (0)} \vert \psi^x_a \ra \vert \downarrow \ra, \eqno
(2.21)
$$
where we use the notation that $\vert \psi^x_a \ra$ contains only
$\alpha ^\mu_n$ oscillators and obeys
$$
L^x_n \vert \psi ^x_a \ra = 0, \quad n\ge 1,\quad (L^x_0 - a)\vert
\psi^x_a\ra = 0.
$$
The intercept $a$ can take only the two values $1$ and $15/16$.
When $a=1$, $\beta$ can have the values  ${6iQ/7}$ or
${8iQ/7}$, and when $a$ is 15/16 $\beta$ takes the value $iQ$.
In reference [18] the spectrum generating algebra for
these states was found and used to classify them.  It was shown that the
count of states was described by Ising model characters $\chi_{h}$
where $h = 1 - a$.

A few examples of physical states that were of non-standard type
were found in references [14,16].  It also emerged, however, that
the $W_3$ string must contain more than the states of standard ghost
type to be consistent.  It was shown that the scattering of four
states of intercept $15/16$ led to an infinite number of intermediate
states with intercept $1/2$ [17], and that modular invariance
required states whose count was given by the Ising character
$\chi_{1/2}$ [18].
It was further pointed out [18]
that  states of the form $\vert\psi ^x_{1/2} \ra$ would indeed have
such a count.

By using a vanishing null-state argument, the cohomology of $Q$
was found in reference [18] to contain such
states having the form
$$
\left(d_{-1} + {i\over \sqrt{522}} b_{-1} \right) e^{i
\beta_1
\varphi(0)} \vert \psi^x_{{1/2}}\ra \vert \downarrow \ra,
\eqno(2.22)
$$
where $\beta_1 = {4iQ/7}$.  The same argument also lead to
the states [18]
$$
\left(d_{-1}- {i\over \sqrt{522}} b_{-1} \right) e ^{i\beta_2
\varphi (0)}\vert \psi^x_{{15/16}} \ra \vert \downarrow \ra
\eqno(2.23)
$$
and
$$
\left({2\over 261} b_{-2} + {9\over 522} L_{-1}^{tot} b_{-1} +
W^{tot}_{-1} d_{-1} \right) e^{i\beta_3\varphi (0)} \vert
\psi^x_{1/2} \ra \vert \downarrow \ra, \eqno(2.24)
$$
where $\beta_2 = {3iQ/7}$ and $\beta_3 = {2iQ/7}$.
Thus it became apparent that the cohomology of $Q$ contained copies of
the same states at different ghost number.  Since the theory was
modular invariant with only one copy of states from each of the
sectors $1,{1/2}$ and ${15/16}$, and the cohomology of $Q$
where investigated contained only copies of these states, it was
conjectured [18] that the cohomology of $Q$ consisted of the
states of equations (2.21) and (2.22) corresponding to intercepts
$1,{15/16}$ and ${1/2}$, as well as copies of these states
and possible discrete states.

Using equations (2.13) and (2.14) it is possible to rewrite the
states of equations (2.21) as
$$
\vert 1,0 \ra =  c\,\partial e\, e\, e^{i\beta(1;0)\varphi}
\vert \psi^x_1\ra \vert 0 \ra\eqno(2.25)
$$
$$
\overline {\vert 1, 0 } \ra = c\, \partial e\, e\,
e^{i\overline\beta (1;0)\varphi}
\vert \psi^x_1\ra \vert 0 \ra, \eqno (2.26)
$$
where we have introduced the notation
$$
\beta (1;n) = (8-8n) {iQ\over7}\quad {\rm and}\quad \overline
\beta (1,n) = (6-8n) {iQ\over 7}, \eqno (2.27)
$$
and
$$
\vert {15/16}, 0 \ra = c\, \partial e\, e\,
e^{i\beta({15/16},0)\varphi} \vert \psi^x_{15/16}\ra\vert 0\ra
\eqno(2.28)
$$
where $\beta ({15/16}, n) = ({7-4n}) iQ/7$.

Similarly, up to a constant of proportionality,
the states of equations (2.22), (2.23) and (2.24) become
respectively
$$
\vert {1/2}, 0 \ra = \left ( c\, e - {i \over \sqrt {522}}
\partial e\,e \right) e ^{i\beta ({1/2}; 0 ) \varphi}\vert
\psi^x_{{1/2}} \ra \vert 0 \ra \eqno (2.29)
$$
$$
\vert {15/16}, 1 \ra = \left(c\, e + {i\over \sqrt
{522}}\partial e\, e\right) e^{i\beta ({15/16} ; 1) \varphi}  \vert
\psi^x_{{15/16}} \ra \vert 0\ra \eqno (2.30)$$
and finally
$$
\eqalign{
  \overline{\vert{1/2},0\ra} =
&  (-{4\over 3 \sqrt{58}}b\,c\,\partial e\,e
  -{4\over 3 \sqrt{58}}\partial^2 e\,e \cr
&  + {1\over\sqrt{29}} \partial\varphi\partial e\,e
  + i \sqrt 2 c\,e\,\partial\varphi
  -{3i\over2} c\, \partial e)
  e^{i\overline\beta({1/2},0)\varphi} \vert \psi^{1/2}\ra
  \vert\downarrow \ra,\cr
}\eqno (2.31)
$$
where
$$
\beta ({1/2}, m) = (4-8m){iQ\over 7}\quad{\rm and}\quad
\overline \beta
({1/2}, m) = (2-8m){iQ\over 7}.  \eqno (2.32)$$
The reason for this notation will emerge once we discuss the
cohomology of $Q$ in section 5.

The simplest such states are those that have no $\alpha^\mu_n$
oscillators and so are tachyonic.  In this case
$$
\vert \psi^x_{a} \ra = e^{ip\cdot x} \vert 0,0 \ra,\eqno(2.33)
$$
with ${1 \over2} p\cdot (p-2i\alpha)= a$.  The next simplest possibility is
$$
\vert \psi^x_{a} \ra = \xi\cdot\partial x \,e^{ip\cdot x} \vert
0,0\ra$$
where $p\cdot\xi =0$ and ${1\over2}p\cdot (p-2i\alpha) = a -1$.
We will often write $\vert \psi^x_{a} \ra$ in vertex operator form as
$\vert \psi^x_{a} \ra = V^x(a) \vert 0,0 \ra$. Further low level states
belonging to the cohomology of $Q$ have been found in
references [21] and [22].
\par

{\bf 3. Explicit results of $W_3$ scattering.}
\par
In a recent paper [17] it was shown that the scattering, at tree
level, of $N$ $W_3$ string states is given by
$$
\int {\prod \limits_ i}^\prime dz_i \,V f(z_i).\eqno(3.1)
$$
Here $V$ is the usual scattering vertex in the presence of a
background charge, and $f$ is an Ising model correlation function
that depends on the intercepts of the external states.  To be
specific, if the $N$ external states have effective intercepts
$a_i,\quad i=1, ...,N$, which can take only the values $1$,
${15/ 16}$ or ${1/ 2}$, then $ f = \la \prod \limits^N _{i=1}
\varphi_i (z_i)\ra$ where $\varphi_i$ is the
Ising field of weight $h_i = 1 - a_i$.

This result followed from an application of the group theoretic
approach to string theory.  The
essential steps in this process are the computation of the vertex
$V$ using overlap relations, and then the determination of the
measure $f$ by demanding that null states decouple.  Using this
technique it was possible to work with the reduced subspace of the
full $W_3$ Fock space in which the physical states sit.  It is
important to understand that the properties of the vertices and
null states used in this reduced Hilbert space are those inherited
from the full Fock space of the $W_3$ string.
\par
We found that the decoupling of the null states of the $W_3$ string
implied that $f$ obeyed the differential equations satisfied by the
Ising model correlators, which are
$$
{4\over 3}{\partial^2 f\over \partial z^2_j} - \sum \limits
^N _
{{i =1\atop i \not= j}} \left\{ { \partial f \over \partial z^i} {1
\over (z^j - z^i)} + {1\over 16} {f\over (z^j - z^i)^2}\right\} = 0,
\eqno (3.2)
$$
$$
{3\over 4}{\partial^2 f\over \partial z^2_j} - \sum \limits
^N _{{i
= 1 \atop i \not= j}} \left \{ {\partial f \over \partial z^i} {1
\over (z^j - z^i)} + {1\over 2} {f \over ( z^j - z^i)^2} \right \}
=
0 \eqno (3.3)
$$
and
$$
{\partial f \over \partial z^j} = 0, \eqno (3.4)
$$
when the $j^{th}$ leg or string has intercept ${15/16}$,
${1/2}$ and $1$ respectively.
\par
The measure $f$ must also satisfy the usual $sl(2,\bb{R})$ relations
$$
\sum \limits_i {\partial\over\partial z^i} f  =
\sum \limits_i (z_i {\partial\over\partial z^i} +
h_i) f
= ( \sum \limits_i z^2_i {\partial\over\partial z^i} +
2h_i z_i ) f =0 \eqno (3.5)$$
as a result of the corresponding relations for the vertex of
equation (3.1) (see equations (9) and (10) of reference [17]).
\par
We refer the reader to reference [17] for further details and in
particular for a discussion of how the two solutions to the
above equations must be
combined to ensure duality.

It is straightforward to evaluate the $W_3$ string scattering
given by equation (3.1) whenever the Ising model correlation
functions, or equivalently the solutions to the above differential
equations, are known.  In reference [1] we carried this out for
certain cases.  In this paper we extend these results so that they
can be compared with the covariant approach to $W_3$ string
scattering.

Clearly, when a leg is from the intercept 1 sector, the measure $f$
does not depend upon that Koba-Nielsen coordinate.  Consequently,
the form of $f$ is the same as that for the scattering process with
only states of intercept ${15/ 16}$ and ${1/ 2}$.  Having
found $f$ it is straightforward to evaluate the scattering for any
external physical states, but in order to be concrete we will often
evaluate the scattering of tachyon states. Applying such states to the
vertex $V$ leads to the expression
$$
\prod \limits _{i<j} ( z^i - z^j )^{2\alpha^\prime p_i \cdot p_j}
\eqno (3.6)
$$
For four tachyon scattering, with the choice of Koba-Nielsen
coordinates $z_1 = \infty$, $z_2 =1$, $z_3 = x$ and $z_4 = 0$,
this reduces to
$$
x^{-\alpha^\prime s - a_3 - a_4} (1 - x)^{-\alpha^\prime
t - a_2 -a_3} ,\eqno (3.7)
$$
where $s = -(p_1 + p_2)\cdot(p_1 + p_2 - 2{i\alpha})$,
$t = -(p_2 + p_3) \cdot(p_2 + p_3 - 2{i\alpha})$ and  a
factor of $(z_1)^{-2a_1}$,
which is cancelled by other such factors, has been removed.

In the case of three string scattering the measure $f$ is
determined by the usual $sl(2,\bb{R})$ relations to be of the form
$$
{C \over ( z_1 - z_2 )^{h_1 + h_2 - h_3} ( z_2 -
z_3)^{h_2 + h_3 -h_1} ( z_1 - z_3 )^{h_1 + h_3 -
h_2}} \eqno (3.8)
$$
As is well known, the values that $C$ can take are equivalent to the
fusion rules of the Ising model.

As explained above, the same formula of equation (3.8) can be used
for four $W_3$ string scattering whenever one of the legs has
intercept 1.  We now give some examples of such scattering.  If we
take legs one and two to be intercept ${15/ 16}$ states, leg $3$
an intercept $1$ state and leg $4$ an intercept ${1/2}$ state
then
$$
 f = {1\over (z_1 - z_2)^{- {3/8}}}{1\over (z_2 -
z_4)^{1/2}}{1 \over (z_4 - z_1)^{1/2}}.\eqno (3.9)
$$
Using equation (3.7) we find that such tachyon scattering is given by
$$
\int dx\quad x^{-\alpha^\prime s- {3/2}} (1 - x)^{-
\alpha^\prime
t - {31/16}}. \eqno (3.10)
$$
To give one other example let us consider strings with intercepts
$1, {1/2}, {1/2}$ and $1$.  On the basis of $sl(2,\bb{R})$
invariance,
$$
 f = {1\over (z_2 -z_3)}, \eqno (3.11)
$$
and tachyon scattering is given by
$$
\int d x \quad x^{-\alpha^\prime s - {3/2}} (1-x)^{-\alpha
^\prime t-2}. \eqno (3.12)
$$

We now consider the scattering of strings all having the same
intercept.  If the intercept is $1$ then $f=1$, and the scattering
is the same as in the ordinary bosonic string with a background
charge [17].  The Ising correlation function of $2N$ weight $1/2$
states is well known [24] and hence [17]
$$
f = Pf {1\over z_{ij}} \equiv {1\over 2^N N!} \sum \limits
_\sigma (-1)^\sigma {1\over z_{\sigma_1 \sigma_2}}\ldots {1\over
z_{\sigma_{2N-1}\sigma_{2N}}} \eqno (3.13)$$
where $z_{ij} = z_i - z_j$.  For $2N = 4$ this becomes
$$
f = \left ({1\over z_{12} z_{34}} - {1\over z_{13}z_{24}} + {1\over
z_{14}z_{23}} \right ). \eqno (3.14)$$
Using equation (3.7) we find that four tachyon intercept ${1/2}$
strings scatter according to
$$
\int  dx \, x^{-\alpha^\prime s-2} (1-x)^{-\alpha^\prime t
- 2} (1-x+ x^2), \eqno (3.15)$$
while for $2N$ such tachyonic states we obtain
$$
\int {\prod \limits _i}^\prime d z_i \prod \limits _{i<j} (z_i -
z_j)^{2\alpha ^\prime p_i\cdot p_j} Pf {1\over z_{ij}}. \eqno (3.16)
$$
The $\prod^\prime _i$ means we omit three indices, say $k,l,m$, from the
product, but instead include the factor $z_{kl} z_{km} z_{lm}$.
\par
Let us now consider $2N$ intercept ${15/16}$ strings.  The
correlation function for $2N$ weight ${1/16}$ Ising states can
be computed from the Feigin-Fuchs construction [20].  Given a
scalar field $\phi$ with an energy momentum tension $T= - {1\over
2} (\partial \phi)^2 - \alpha _0 \partial ^2 \phi$, the central
charge is $c = 1 + 12 \alpha^2_0$.  Hence if $c={1/ 2}$ we
require $ \alpha _0 = {i/2\sqrt6}$.  The Ising fields of
weights $0, {1/2}$ and ${1/16}$ are represented by the
fields $V_\alpha =  e^{i\alpha \phi}$ where $\alpha$ takes the values $0,
-4 i \alpha_0, 3i\alpha_0$ (or $2i\alpha_0 -
\alpha$) respectively.  As is well known [20] one requires
screening charges in order to obtain the correct Ising model
correlation functions since one cannot balance the momentum to be $
2i\alpha_0$.  These screening charges are given by $S_\pm = \oint d
 w \,e^{i\alpha_\pm \varphi}$, where ${1\over 2}\alpha_\pm
(\alpha_\pm - 2 i\alpha_0) = 1$, and so  $ \alpha_+ = -6i
\alpha_0, \quad \alpha_{-}= 8i\alpha_0$.  For $2N$ weight $1/16$
fields we can achieve such a balance by taking the fields
$$
\la 0 \vert V_{2i\alpha_0 - \alpha}(z_1) V_\alpha(z_2) V_\alpha(z_3)
\ldots V_\alpha (z_{2N})  S_+^{N-1}
\vert 0 \ra ,\eqno (3.17)
$$
which results in the correlation function
$$
\la\prod\limits ^{2N}_{j=1} \sigma (z_j) \ra =
\int\left(\prod^{N-1}_{i=1} dw_i\right)
\la e^{i(2i\alpha_0-\alpha) \phi (z_1)} \prod
\limits^{2N}_{j=2} e ^{i\alpha \phi (z_j)} \prod\limits^{N-1}_{k=1}
e^{i\alpha_+ \phi (w_k)}\ra \eqno (3.18)$$
$$\eqalign{
= \int\left(\prod^{N-1}_{i=1} dw_i\right)
  \prod \limits^{2N}_{j=2} (z_1 -& z_j)^{-{1/8}} \prod
  \limits^{N-1} _{k=1} (z_1 -w_k)^{1/4} \prod
  \limits^{2N}_{{i<j\atop i,j=2}} (z^i - z^j ) ^{3/8}\cr
& \prod\limits _{{k,l\atop k<l}} (w_k -w_l)^{3/2}
  \prod \limits_{i,k} (z_i - w_k)^{-{3/4}}.\cr}
$$
For 4 such states this becomes
$$
 \int dw\, (z_1 - w)^{1/4} \prod \limits^4_{i=2}(z_1-z_i)^{-
{1/8}} \prod \limits^4 _{i>j} (z_i-z_j)^{3/8} \prod
\limits^4_{i=2} (z_i-w)^{-{3/4}}, \eqno (3.19)
$$
which, using the canonical choice for the $z_i$, becomes
$$
\eqalign{&\left(x(1-x)\right)^{{3/8}} \int dw \left(w (1-
w)(x-w)\right)^{-{3/4}}\cr
& = \left((1-x)x\right)^{{3/8}} \left[a\,F
({3/4}, {5/4}, {3/2}; z) + b\, z^{-{1/2}}
F({3/4}, {1/4}, {1/2}; z) \right]\cr}\eqno(3.20)$$
where $a$ and $b$ are constants and where we have removed a factor
of $z_1^{1/8}$ that cancels with other such factors elsewhere.
Using results from appendix A we find the result for the correlation
function of four weight $1/16$ fields to be
$$
{1\over \left(x(1-x)\right)^{1/8}} (a^\prime \sin{\theta\over2}
 + b \cos{\theta \over2}), \eqno(3.21)
$$
where $x =   \sin^2\theta$.  This is the well known result [24].
An extensive discussion of four $W_3$ tachyonic intercept ${15/16}$
scattering was given in reference [17]\footnote{$\dagger$}{There is an
obvious misprint in
equations (19) and (20) of reference [17], but the subsequent
equations are recorded correctly.}.

As we shall see the Feigin Fuchs construction provides
a useful model, some of whose features we will exploit in the
covariant formulation of $W_3$ scattering.
One could continue to compute more examples of $W_3$ string
scattering, but the above will more than suffice.

{\bf 4. Screening charges and the scattering of 4
intercept 15/16\ $W_3$ strings}

In this section we wish to give an alternative construction of
$W_3$ string scattering to that of reference [17].  We intend to
mimic for the $W_3$ string what has become known as the conformal
field theory approach to string theory, which involves constructing
amplitudes from the ghost fields and vertex operators.  This
generalizes the original approach to string scattering which used
only the latter objects.  These procedures [19] for the usual bosonic
and superstring theories are well known and can act as a useful
guide in formulating a set of rules.  We must also ensure that the
amplitudes are BRST invariant, which corresponds in the old
method to the decoupling of null states.  However, given a new
theory such as $W_3$ string theory one must essentially guess a
good set of rules and then check the consistency of the amplitudes
with themselves and with the same amplitudes found in other
formalisms.

The vertex operators can be read off from the states that occur in
the cohomology of Q.  As was explained in Section 2 such states are
associated with one of three intercepts $1, {1/2}$ and
${15/16}$.  Using the notation employed in section 2 for the states,
we find the following vertex operators:
$$
\eqalignno{V(1,0)& = c\,\partial e\, e\
e^{i\beta(1;0)\varphi}V^x(1)&(4.1)\cr
\bar V(1,0)& =\ c\,\partial e\, e\
e^{i\bar\beta(1;0)\varphi}V^x(1),&(4.2)\cr}
$$
$$
\eqalignno{V({1/2},0) &= \biggl(c\,e-{i\over \sqrt{522}} \partial
e\,e\biggl)\ e^{i\beta({1/2},0)\varphi}V^x({1/2})&(4.4)\cr
\bar V({1/2},0) &= \biggr(
-{4\over 3 \sqrt{58}}b\,c\,\partial e\,e
  -{4\over 3 \sqrt{58}}\partial^2 e\,e \cr
&  + {1\over\sqrt{29}} \partial\varphi\partial e\,e
  + i \sqrt 2 c\,e\,\partial\varphi
  -{3i\over2} c \partial e\biggl)
e^{i\bar\beta({1/2},0)\varphi}V^x({1/2}),&(4.5)\cr}
$$
and
$$
\eqalignno{V({15/16},0)& =\ c\partial e\ e\
e^{i\beta({15/16},0)\varphi}V^x({15/16})&(4.6)\cr
V({15/16},1)& =\ \biggl(c\,e+{i\partial
e\,e\over\sqrt{522}}\biggr)\ e^{i\beta({15/16},1)\varphi}
V^x({15/16}).&(4.7)\cr}
$$
Here $V^x(a)$ is any vertex operator, constructed from $x^\mu$
alone, that has conformal weight $a$ with respect to $T^x(z)$.  The
simplest is $V^x(a)=e^{ip\cdot x}$, where ${1\over 2}p\cdot(p-2i\alpha)=a$.
It is straightforward if tedious to verify that Q
commutes with all these vertex operators,  as expected.

Given the states or vertex operators we may find new states or
vertex operators by acting with the operators
$[\varphi,Q]$
and
$[x^\mu,Q]$.
It is well known that, although these operators are formed from
commutators with Q, they do not lead
to BRST trivial operators when acting on vertex operators,
since $\varphi$ and $x^\mu$ are not
well-defined conformal fields.  Extensive use has been made of
these operators in two dimensional gravity theories [23] and in
the analogue of these states in the two scalar $W_3$ string [16].
They
were also used in references [21] and [22].  Such operators
also occur in the superstring, but there $\varphi$ and $x^\mu$ are
replaced by one of the bosonized ghost fields.  There they lead to
the phenomenon of picture changing [19], and in view of the
similar role in the played by these operators in the
$W_3$ string we shall refer to them here as
picture changing operators.

It is convenient to work with
some linear
combination $P$ of $[x^\mu,Q]$ and $[\varphi,Q]$, chosen
such that the action of
$P$ on a vertex operator having the form of a function of ghosts and
$\partial\varphi$ multiplied by $V^x(a)$ leads to another vertex
operator of the same form.
Since the  operator $P$ commutes with $Q$,
the product $P(z)V(w)$ will also commute with $Q$ if $V(w)$ does.
Consequently all terms
in the operator product expansion, including the term independent
of $z$,
$$
:PV:(w)= \oint\limits_w dz\ {P(z)V(w)\over (z-
w)},\eqno(4.10)
$$
will also commute with $Q$.

One could attempt to use the above vertex operators alone to
construct string scattering amplitudes.  Such an approach would be
in complete analogy with other well understood string theories.
This was the approach adopted in references [21,22].  However, in
the $W_3$ string we have a background charge, and the vertex
operators can possess only certain values of the $\varphi$ momentum.
These values are discussed more fully in the next section.  It is readily
apparent that one is unable, in general, to use the vertex operators alone in
such a way that their total momentum adds up to the required value $2iQ$.
This is particularly apparent for $2N$ intercept ${15/16}$
states, since the allowed $\varphi$ momenta are
$\beta({15/16},n)=(7-4n){iQ/7},\ n \in Z_+$,
which would require the
condition $-2\sum\limits_i n_i=7(N-2)$ with $n_i \in \bb{Z}_+$.
In reference [17], however, these amplitudes were found not to vanish,
and in section 5 we shall give a physical argument for why they must be
non-vanishing.

The way out of this apparent paradox is to use screening charges,
that is, objects of the form
$$\int dw\ e^{i\beta\varphi}f(b,c,d,e,\partial\varphi)\eqno(4.11)$$
that commute with $Q$.
If we can find such charges that do not involve the field $x^\mu$,
then the insertion of these charges into a correlation function will not
change the effective space-time interpretation of the correlation
function.

Such charges will commute with $Q$ only if the integrand has
weight 1, which, since $f$ is a function of the ghosts and
$\partial\varphi$,
means that ${1\over 2}\beta(\beta-2iQ)=n$ for  $n \in \bb{Z}$.
If we write $\beta={isQ/7}$ then $s^2-14s-16n=0$, which implies that
$$s=7\pm\sqrt{49-16n}.\eqno(4.12)$$
For s to be an integer, as it appears always to be, we must have
$49-16n=m^2,\,m \in \bb{Z}$.
This can only be the case if $m=8N\pm 1$
and $n=-(4N\mp3)(N\pm 1)$, with $N \in \bb{Z}$.  The values of $n$ so found are
$3,\ 0,\ -2,\ -11,\ -15,\ldots $.  The function $f$ must have weight
$1-n$, that is $-2,\ 1,\ 3,\ 12,\ 16,...$.  There are some
obvious candidates for weights $-2$ and $3$,
namely $f=e+\dots$ and $f=d+\dots$ respectively.

To evaluate the commutator of $Q$ with the screening charge, we use
the formula
$$[Q,\ (fe^{i\beta\varphi})(z)]\ =\ \oint\limits_z dw\,
j^{BRST}(w)(fe^{i\beta\varphi})(z)\eqno(4.13)
$$

We first consider $f = e$, in which case we need the formula
$$
\eqalign{[&Q,e\ e^{i\beta\varphi}(z)]\ =\ \partial[c\ e\
e^{i\beta\varphi}]\cr +
&{i\over 6}\beta(\beta-iQ)(\beta-2iQ)
\partial[\partial e\,e\, e^{i\beta\varphi}]\cr +&\beta(\beta-iQ)\ {\partial
e\ e\partial\varphi\over 12}[{1\over 2}\beta(\beta-2iQ)-
3]\cr}\eqno(4.14)
$$
In the derivation of this we used the result
$$
\eqalign{
W^m(z)\ e^{i\beta\varphi}(w)&={i\over 3}{\beta(\beta-2iQ)(\beta-iQ)
\over (z-w)^3}e^{i\beta\varphi}(w)\cr
-&{\beta(\beta-
iQ)\over (z-w)^2} \partial\varphi\ e^{i\beta\varphi}(w)\cr
& +\ {1\over z-w}(-i\beta(\partial\varphi)^2
-2\beta T^x-
\beta^2\partial^2\varphi)\ e^{i\beta\varphi}(w).\cr}\eqno(4.15)
$$
In the case we are considering,
$n\equiv {1\over 2}\beta(\beta-2iQ)=3$, and so $\beta$
takes the values $\bar\beta^s_1={6iQ/7}$ or
$\beta^s_1={8iQ/7}$.  Then the last term of
equation (4.14) vanishes, and it is trivial to see that
$$
[Q,\int dz\, e\,e^{i\beta^s_1\varphi}(z)]=0=[Q,\int dz\,e\,
e^{i\bar\beta^s_1\varphi}(z)]\eqno(4.16)
$$

The next interesting possibility is $n\equiv {1\over 2}\beta(\beta-
2iQ)=-2$, so that $\beta=-{2iQ/7}$ or ${16iQ/7}$.  Taking
$f$ to have terms of ghost number $-1$ and weight $3$ one finds after a
lengthy calculation that
$$
[S,Q]=0
$$
where
$$
S = \oint dz \{d-{5i\over 3\sqrt{58}} \partial b-{2\over
3.87} \partial b\,b\,e-{4i\over 3}{1\over \sqrt{58}} d\,b\,e
\}e^{i\beta^s\varphi}\eqno(4.17)
$$
and $\beta^s=-{2iQ/7}$.

We are now in a position to construct $W_3$ string scattering, with
the building blocks being the vertices and the screening charges
above.  It is possible to construct all amplitudes using only the
vertices $V(1,0)$, $V({1/2},0)$ and $V({15/16},0)$ together with the
picture changing operator $P$ and the screening charge $S$.  The
reason for this will become apparent in  section 6.  For our
initial example, however, we will use the additional vertex
operator $V(15/16,1)$ [18].

A useful standard man\oe uvre for reducing the number of $c$ ghosts
contained in a vertex is to realise that if $V(z)$ is any vertex
that commutes with $Q$ then
$$
\int dz\,\oint\limits_z dv\, b(v)\, V(z)\eqno(4.18)
$$
also commutes with $Q$.  This follows from the relation
$$
[Q,[\oint\limits_z\ dv\ b(v)\ V(z)]]\ =\
\oint\limits_z\ dv\ T^{tot}(v)\ V(z)\ =\ [L^{tot}_{-1}, V(z)]\ =\
\partial_zV(z).\eqno(4.19)
$$

To illustrate this procedure, we compute the scattering of 4
sector ${15/16}$ states.  We can balance the $\varphi$
momentum by taking one vertex operator $V({15/16},0)$  with
momentum $\beta({15/16},0)={7iQ/7}$, three vertices
$V({15/16},1)$ with momentum $\beta({15/16},1)={3iQ/7}$,
and one screening charge $S$ with momentum $\beta^s=-{2iQ/7}$,
since then
$$
\beta({15/16},0) + 3\beta({15/16},1) + \beta^s =
2iQ.\eqno(4.20)
$$
To obtain the correct number of $c$ ghosts, namely 3, we make use
of equation
(4.18) applied to one of the $V({15/16},1)$, i.e.
$$
\int dz\oint\limits_z dv\, b(v)\ V({15/16},1)(z)\ =\int dz\, e\,
e^{i\beta({15/16},1)\varphi}V^x({15/16})(z).\eqno(4.21)
$$
We also require 5 factors of $e$ and so we must apply one picture
changing operator to, say, $V({15/16},0)$.  This gives
$$
PV({15/16},0)\ =\ c\,\partial^2e\, \partial e\, e\,
e^{i\beta({15/16},0)\varphi}\ V^x({15/16})\eqno(4.22)$$

The amplitude for 4 intercept ${15/16}W_3$ string scattering
is thus given by
$$
\eqalign{
\la 0 \vert P & V(15/16,0)(z_1) V( 15/16,1)(z_2)\cr
&  \oint dz_3 \oint_{z_3}dv\,b(v)
  V(15/16,1)(z_3) V(15/16,1)(z_4) S \vert 0 \ra\cr
= & \oint dz_3 \int dw\, \la 0 \vert
  \left( c\,\partial^2 e\,\partial e\, e\ e^{i\beta(15/16,0)\varphi}
  V^x(15/16)\right)(z_1)\cr
& \left( c\,e\ e^{i\beta(15/16,1)\varphi}V^x(15/16)\right)(z_2)
 \left( e\ e^{i\beta(15/16,1)\varphi}V^x(15/16)\right)(z_3)\cr
& \left( c\,e\ e^{i\beta(15/16,1)\varphi}V^x(15/16)\right)(z_4)\cr
&  \left((
d-{5i\over 3\sqrt{58}} \partial b-{2\over
3.87} \partial b\,b\,e-{4i\over 3}{1\over \sqrt{58}} d\,b\,e
)e^{i \beta^s\varphi}\right)(w) \vert 0 \ra\cr
= & - \oint dz_3 \int dw \la 0 \vert
  \left(\prod_{i=1\atop i\ne 3}^4 c(z_i)\right)
  \left(\partial^2 e\,\partial e\,e\right)(z_1)
  e(z_2) e(z_3) e(z_4) d(w)\cr
& e^{i \beta(15/16,0)\varphi}(z_1)
  \prod_{i=2}^4 e^{i \beta(15/16,1)\varphi}(z_i)
  e^{i\beta^s \varphi}(w)
  \prod_{i=1}^4 V^x(15/16)(z_i) \vert 0 \ra\cr
}\eqno(4.23)
$$
By bosonizing the ghosts we find the $c$ ghosts give a factor
$$
(z_1-z_2)(z_1-z_3)(z_2-z_3)\eqno(4.24)
$$
and the $e$-$d$ ghosts a factor
$$
(z_1-w)\prod\limits^4_{i=2}(z_1-z_i)^3(z_i-w)^{-
1}\prod\limits^4_{{i,j=2}\atop {i<j}}(z_i-
z_j).\eqno(4.25)
$$
Evaluating the exponential factors in the usual way we find that the
amplitude for tachyonic states is proportional to
$$
\eqalign{&\oint dw\ \int dx\, x(1-x)^{-1/8}[w(1-w)(x-w)]^{-{1\over
4}} (1-x)^{p_2\cdot p_3}x^{p_3\cdot p_4}\cr
=&\int dw\ \int dx\ x^{s/2-2}(1-x)^{t/2-2}[(1-w)(x-w)w]^{-{1\over
4}}\cr}\eqno(4.26)
$$
where we have chosen $z_1=\infty$, $z_2=1$, $z_3=x$ and
$z_4=0$.

Using appendix (A) we find the result
$$
\int dw \int dx\ x^{5/2-2} (1-x)^{t/2-2}\ (a\
\cos\theta/2+b\sin \theta/2)\eqno(4.27)
$$
where a and be are constants and $x = \sin^2\theta$.
This agrees with that found in
reference [17].  This reference also contains detailed
discussion of how crossing allows one to determine the coefficients a
and $b$; the factorization and duality properties of this
result are also discussed.

{\bf 5. The cohomology of Q}

It was discovered in reference [18] that the cohomology of $Q$
contained, in addition to the intercept ${15/16}$ and 1 states of
ghost number zero,  states associated with intercept ${1/2}$
with ghost number $-1$ and also copies of these states
and those with intercept $15/16$ at other
ghost numbers.  It was also shown [18] that in order to build
a modular invariant
theory it was sufficient to take only one copy of each of the states
with intercepts $1$, ${1/2}$ and ${15/16}$.

This leads one to suspect that there may be some simple relations
connecting states in the cohomology of $Q$ associated with the same
intercept.  This is indeed the case;  we will now show that such
states are related by the actions of the screening operator $S$ of
equation (4.17) and the picture changing operator $P$.

Since $S$ commutes with $Q$, it follows that whenever the action of
$S$ on a physical state is well-defined and non-zero it will
produce another physical state.  Before giving a general discussion,
let us consider an example.  We shall show how the vertices
$V({1/2},0)$ and $\bar V({1/2},0)$ can be related by the
action of a single screening charge, after a suitable picture change.
Given only one screening charge $S$ with $\varphi$ momentum $\beta^s$,
its action on a vertex operator
with $\varphi$ momentum $\beta$ will be well defined if
$\beta^s\beta$ is an integer.

For the case under consideration $\beta^s\beta({1/2},0)=1$ and
so the $\varphi$ exponentials combine to provide one power of $(z-
w)$.  Hence, to gain a non-zero result the ghosts must provide at
least a factor of $(z-w)^{-2}$.  A short calculation shows that
although $SV({1/2},0)$ is well defined it vanishes.  We
therefore introduce further powers of the ghost $e$ into the
vertex using a picture changing operator. We have
$$
\eqalign{
PV({1/2},0)= & \bigl( 5 \,\partial^2 e\,e -
  {24 Q\over 7} \partial\varphi \partial e\,e\cr
&  - {19 i\over 3\sqrt{58}} \partial^2 e\,\partial e\,e
\bigr)e^{i\beta(1/2,0)\varphi}V^x(1/2),\cr
}\eqno(5.1)
$$
and acting on this with the screening operator we find the
vertex $\bar V({1/2},0)$,
$$
SPV({1/2},0)\propto\bar V({1/2},0).\eqno(5.2)
$$

Let us now consider when two screening operators have a well
defined action on a vertex operator with $\varphi$ momentum
$\beta$.  Since this has not, to our knowledge, been clearly
discussed in the literature let us consider the more general case
of two screening operators with $\varphi$ momentum $\beta^s_i,
i=1,2$  Since the factors in front of the exponentials are made
from either ghosts or $\partial\varphi$ they can only contribute
integer powers of the coordinates, and consequently do not affect
whether or not the action of the screening operators is well
defined.  Consequently, we must focus our attention on the factors
$$\eqalign{
\oint\limits_{C_1}dw_1\ \oint\limits_{C_2}dw_2\
& e^{i\beta^s_1\varphi(w_1)}\ e^{i\beta^s_2\varphi(w_2)}\
  e^{\beta\varphi(z)}\cr
& = \oint\limits_{C_1}dw_1\
  \oint\limits_{C_2}dw_2\ (w_1-w_2)^{\beta^s_1\beta^s_2}(w_1-
  z)^{\beta^s_1\beta}(w_2-z)^{\beta^s_2\beta}\cr
}\eqno(5.3)
$$
Since $\vert w_1-z\vert \gg \vert w_2-z\vert $ we arrange the
$w_1$ and $w_2$ contours to be around the point $z$ in such a way
that the above condition is satisfied.

Let us consider the substitution $(w_1,w_2)$ to $(w_1,\tau)$ given
by $(w_2-z)=\tau(w_1-z)$.  The above condition implies that
$\vert\tau\vert< 1$, but we also demand that $\tau=1$ at one and
only point in other words the $w_1$ and $w_2$ contours touch one
another.  This latter condition ensures that the value of the
integral is dependent on only the one place where the contours
cross the branch cut.  We are to regard $w_2$ as fixed and consider
the $\tau$ integration.  Substituting $dw_2=d\tau(w_1-z)$ we find
the above integrals become
$$\eqalign{\oint_z dw_1\oint d\tau
&\tau^{\beta^s_1\beta}(1-\tau)^{\beta^s_1\beta^s_2}(w_1-z)^P\cr
&\exp(i\beta^s_1\varphi(z+(w_1-z))+
i\beta^s_2\varphi(z+\tau(w_1-z))+i\beta\varphi(z))\cr}\eqno(5.4)
$$
where $P=1+\beta^s_1\beta+\beta^s_1\beta^s_2+\beta^s_2\beta$.  This
integral is well defined if $P$ is an integer.

The generalization
to $n$ screening charges with $\varphi$ momenta $\beta^s_i$ is
straightforward.  Their action contains the term
$$
\eqalign{
\prod\limits^n_{i=1}\left(\oint dw_i\,
  e^{i\beta^s_i(w_i)}\right) e^{i\beta\varphi(z)} = & \cr
\oint
  dw_i\prod\limits^n_{{i<j}\atop {i,j=1}} (w_i-
  w_j)^{\beta^s_i\beta^s_j}\prod\limits^n_{j=1} (w_j-
  z)^{\beta^s_j\beta}
& \exp i\{\sum\limits^n_{j=1}\beta^s_j\varphi(w_j)+\beta\varphi(z)
  \}\cr
}\eqno(5.5)
$$
We now replace $w_j$, $j=2,\dots,n$ by $w_1,\tau_j\ j=2,...,n$
using the formula $(w_j-z)=\tau_j(w_1-z)$.  The above expression
becomes
$$
\oint\limits_z\ dw_1\prod\limits^{n-1}_{i=1}\oint d\tau_i\
f(\tau_i)\ (w_1-z)^{P^\prime}\ \exp
i\left\{\sum\limits^n_{j=1}\beta^s_j\varphi(z+\tau_j(w_n-
z))+\beta\varphi(z)\right\}\eqno(5.6)
$$
where $f(\tau_i)$ is a function of $\tau_i$ and
$$
P^\prime\ =\ (n-1)+\sum\limits^N_{{i,j=1}\atop{i<j}}\
\beta^s_i\beta^s_j + \sum_{j=1}^N \beta^s_j\beta.\eqno(5.7)
$$
Thus the integrals are well defined if $P^\prime$ is an integer.

Let us apply this general discussion to the case of interest,
namely the action of $n$ screening charges $S$ with momentum
$\beta^s=-{2iQ/7}$.  Taking $\beta={isQ/7}$ with $s$ an integer, their
action is well defined if
$$
{n\over 4}[-(n-1)+s] \in \bb{Z},\eqno(5.8)
$$
and then the momentum of the resulting vertex is
$$
{iQ\over 7}(s-2n).\eqno(5.9)
$$
For the action on $V({15/16},0)$ we have $s=7$. We then find that $n$
must be even and that the vertices so constructed have momenta
$$
\beta({15/16},m)\ =\ {iQ\over 7}(7-4m),\quad
m \in \bb{Z}_+ .\eqno(5.10)
$$
For the action on $V(1,0)$, for which $s=8$, we find that $n=4m$ or
$4m+1$ with $m \in \bb{Z}_+$, which leads to the vertices with
momenta
$$
\beta(1,m) = (8-8m){iQ\over 7},\quad
m \in \bb{Z}_+,\eqno(5.11)
$$
and
$$
\bar\beta(1,m) = (6-8m){iQ\over 7}\quad
m \in \bb{Z}_+.\eqno(5.12)
$$
We can also consider the action on $\bar V(1,0)$, but in this case
we find the
same series of momenta.

Finally the action of $n$
screening charges on
$V({1/2},0)$ is well defined if $n=4m$ or $4m+1$ with
$m \in \bb{Z}_+$, which leads to vertices with momenta
$$
\beta({1/2},m)\ =\ {iQ\over 7}(4-8m)\eqno(5.13)
$$
and
$$
\bar\beta({1/2},m)\ =\ {iQ\over 7}(2-8m).\eqno(5.14)
$$

It can and does happen that applying $S$'s alone leads to a
vanishing result.  This can be avoided, however, by the judicious
use of the picture changing operator $P$.  Before giving a general
discussion of this point we give another example.  Let us consider the
intercept ${15/16}$ vertices and show how to go from
$V({15/16},0)$ to $V({15/16},1)$.  As is clear from the above,
we require two screening charges.  However, their action
will vanish unless we first act with a picture changing operator.
We find that
$$
PV({15/16},0)\ \propto\ c\,\partial^2e\, \partial e\,e\
e^{i\beta({15/16},0)\varphi}V^x({15/16}),\eqno(5.15)
$$
and acting with $S^2$ we find
$$
\eqalign{
S^2(PV({15/16},0))=&\oint\ dw_2\ \oint\ dw_1\cr
& (d-{5i\over 3\sqrt{58}} \partial b-{2\over
  3.87} \partial b\,b\,e-{4i\over 3}{1\over \sqrt{58}} d\,b\,e)(w_1)
  e^{i\beta^s\varphi(w_1)}\cr
& (d-{5i\over 3\sqrt{58}} \partial b-{2\over
3.87} \partial b\,b\,e-{4i\over 3}{1\over \sqrt{58}}
d\,b\,e)(w_2)e^{i\beta^s\varphi(w_2)}\cr
&(c\,\partial^2e\ ,\partial e\,e\ e^{i\beta({15/16},0)\varphi}\
e^{ip\cdot x})(z)\ V^x({15/16})\cr
}\eqno(5.16)
$$
Bosonizing the ghosts,the term with two $d$'s is proportional to
$$
\eqalign{
& c(z)\ V^x({15/16})(z)\cr &\oint dw_2\
\oint dw_1 e^{-i\rho(w_1)} e^{-i\rho(w_2)} e^{i\beta^s\varphi(w_1)}
e^{i\beta^s\varphi(w_2)}e^{3i\rho(z)}
e^{i\beta({15/16},0)\varphi(z)}\cr =&c(z)V^x({15/16})(z)\
\oint dw_2\oint
dw_1(w_1-z)^{-5/4}(w_2-z)^{5/4}\cr &(w_1-w_2)^{1/2}\exp\
i\{-\rho(w_2)+3\rho(z)\}\cr &\exp\
i(\beta^s\varphi(w_1)+\beta^s\varphi(w_2)+\beta({15/16},0)
\varphi(z))\cr}\eqno(5.17)
$$
We now make the substitution $w_1-z=\tau(w_2-z)\ ,\
\vert\tau\vert<1$, to find that the above becomes
$$
\eqalign{&S^2(PV({15/16},0)) \propto c(z)\ V({15/16})(z)
\oint\limits_z dw_2\int d\tau\ \tau^{-5/4}\cr &(\tau-
1)^{1/2}(w_2-z)^{-1}\ \exp\ i\ (-\rho(z+\tau(w_2-z))\cr &-
\rho(z+(w_2-z))+3\rho(z))\exp\ i(\beta^s\varphi(z+\tau(w_2-z))\cr
&+\beta^s\varphi(z+w_2-z)+\beta({15/16},0)\varphi(z))\cr
&=c(z)V^x({15/16})(z)(e\ e^{i\beta({15/16},1)\varphi}(z)
\int d\tau\ \tau^{-5/4}(\tau-1)^{1/2}\cr
&\propto\ c\ e\ e^{i\beta({15/16},1)\varphi}
V^x({15/16})\cr}\eqno(5.18)
$$
which we recognise as the first term in the
vertex $V({15/16},1)$.
The additional terms are calculated in a similar way.  For example,
the term with $d(w_1) (-5i/(3\sqrt{58})\,\partial b(w_2))$ provides
us with a ghost contribution of $-\tau^{-3}(w_2-z)^5$, and so leads
to a contribution $ (-5i/(2.3\sqrt{58}))\partial e\,e$ multiplied
by exponentials.  Calculating the three other terms one finds
$$
V({15/16},1)\ =\ S^2(PV({15/16},0)).\eqno(5.19)
$$
It is straightforward, if tiresome, to show that we can repeat this
operation to find the vertex $V({15/16},2)$, which is given by
$$
V({15/16},2)\ =\ S^2\ P\ V({15/16},1)\ =\
S^2PS^2PV({15/16},0)\eqno(5.20)
$$
In this way the general pattern for the vertices in the sector $15/16$
emerges.  We have the vertices
$$
V({15/16},m)\ =\ (S^2P)^m\ V({15/16},0)\eqno(5.21)
$$
and in addition
$$
P\ V({15/16},m).\eqno(5.22)
$$

Let us turn our attention to the intercept $1$ vertices.
Although $S$ gives zero when acting on $V(1,0)$, we can act with $P$
first to find the vertex
$$
P\ V(1,0)\ =\ c\, \partial^2 e\, \partial e\, e\
e^{i\beta(1,0)}V^x(1).\eqno(5.23)
$$
The action of $S$ on this vertex does give a non-zero result,
namely
$$
\bar V(1,0)\ =\ c\,\partial e\ e\ e^{i\bar\beta(1,0)}V^x(1)\ =\
SPV(1,0).\eqno(5.24)
$$
To find the next vertex $V(1,1)$ we must act with $P$ and then
three screening charges.  Since this provides us with a simple
example with three screening charges we will give a few of the
details.  We act on the state
$$
P\bar V(1,0)\ =\ c\,\partial^2e\ e\
e^{i\bar\beta(1,0)\varphi}V^x(1)\eqno(5.25)
$$
with three screening charges
$$
\prod\limits^3_{i=1}\left(\int
dw_i
(d-{5i\over 3\sqrt{58}} \partial b-{2\over
3.87} \partial b\,b\,e-{4i\over 3}{1\over \sqrt{58}} d\,b\,e)
e^{i\beta^s\varphi(w_i)}\right) P\bar V(1,0).
$$
The leading term with three $d$'s is given by
$$
\eqalign{
&=\ cV^x(1)(z)\prod\limits^3_{i=1}
\oint dw_i\prod^3_{i=1}(e^{-
\rho(w_i)})e^{3i\rho(z)}e^{i\bar\beta(1,0)\varphi(z)}\cr &=\ c\
V^x(1)(z)\prod\limits_i\oint dw_i\prod\limits_{i<j}(w_i-
w_j)^{1/2}\prod\limits^3_{j=1}(w_j-z)^{-3/2}\cr &\exp\
i\biggl(-\sum\limits_j\rho(w_j)+3\rho(z)\biggr)\ \exp\
i\biggl(\sum\limits_i\beta^s\varphi(w_j)+\beta\varphi(z)\biggr)\cr
&=\ c\ V^x(1)(z)\oint d\tau_1\oint d\tau_2[(\tau_1-\tau_2)(\tau_1-
1)(\tau_2-1)]^{1/2}\cr &(\tau_1-\tau_2)^{-3/2}\ exp\
i(\beta(1,1)\varphi(z))\cr &= c\
e^{i\beta(1,1)\varphi(z)}V^x(1)(z).\cr}\eqno(5.26)
$$
Evaluating the remaining terms we find the result
$$
\left(
c + {7i\over 3\sqrt{58}}\partial e - {8\over 261}b\,\partial e\,e
- {4i\over 3 \sqrt{29}}\partial\varphi\,e\right)e^{i\beta(1,1)\varphi}V^x(1),
$$
which can be written as
$$
V(1,1)\ =\ S^3P\bar V(1,0)\ =\ S^3PSPV(1,0).\eqno(5.27)
$$

It is straightforward to compute further vertices, but the general
pattern now emerges.  We have the vertices
$$
V(1,m)\quad {\rm and}\quad \bar V(1,m)\eqno(5.28)
$$
as well as
$$
PV(1,m)\quad {\rm and} \quad P\bar V(1,m).\eqno(5.29)
$$
These vertices are defined by the relations
$$
\bar V(1,m)\ =\ SPV(1,m),\eqno(5.30)
$$
$$
V(1,m)\ =\ S^3P\bar V(1,m-1).\eqno(5.31)
$$

The vertices with intercept ${1/2}$ have a
similar pattern to those of intercept 1.  Starting from the vertex
$V({1/2},0)$ we can, by acting with $P$ and $S$, create the
vertices
$$
V({1/2},m)\quad {\rm and}\quad \bar V({1/2},m)\eqno(5.32)
$$
as well as
$$
P\ V({1/2},m) \quad{\rm and}\quad P\bar V({1/2},m)\eqno(5.33)
$$
by using the relations
$$
V({1/2},m)\ =\ S^3P\bar V({1/2},m-1)\eqno(5.34)
$$
and
$$
\bar V({1/2},m)\ =\ SPV({1/2},m).\eqno(5.35)
$$

To summarise, we have found that given the basic vertices
$V(a,0)$ for $a=1$, ${1/2}$ and ${15/16}$, we can use $S$ and
$P$ to create the BRST invariant vertices $V(a,m)$; for $a=1,1/2$
we obtain in addition the vertices $\bar V(a,m)$.
We can also obtain further vertices by the action of $P$
on these.  Leaving aside the question of discrete states and the
further action of picture changing operators it would seem most
likely that these are the only states in the cohomology of $Q$,
thus extending the spectrum conjecture of reference [18].
We have now amassed
considerable evidence for this extended conjecture.  One
copy of the states from each sector is known to be sufficient for
modular invariance [18] and is consistent with factorization of
tree level scattering [17].  Further, any states that arise
with this effective intercept can be of positive norm only for the
intercept vertices $1,{1/2}$ and ${15/16}$ [18].  These facts
make it rather unlikely that the cohomology of $Q$ could contain
states not of the above type.  Finally, the vertices above lead to
all vertices known to belong to the cohomology of $Q$ and in
particular to the $V({1/2},0)$, $\bar V({1/2},1)$and
$V({15/16},1)$ found in reference [18] and the additional vertices
found in references [21,22].

It was shown in reference [18] that the physical states for a
given intercept have a spectrum generating algebra involving the
operators $B^i_n,i=1,\dots D-2$ and $B_n$.
Consequently, the conclusion given above
is that states in the cohomology of $Q$ are generated by
$B^i_n,B_n,S$ and $P$.

{\bf 6. General Formalism for $W_3$ String
Scattering}

The $W_3$ string scattering amplitudes are to be constructed
from the building blocks
$$
V(a,0)\ ;\ a=1,{15/16},{1/2}\ ;\ S\ ,\ P\eqno(6.1)
$$
and the operation
$$
\int dz\oint\limits_z dv\, b(v).\eqno(6.2)
$$

The latter is a standard operation used for the $b-c$ ghost system.
We must, however, assemble the building blocks so that the $\varphi$
momentum sums to $2iQ$.  This tells us the required number $N_s$ of
screening charges.  We must also have, after carrying out all the
operator product expansions, $3\ c$ ghosts and $5\ e$ ghosts,
otherwise the correlator will vanish.  As usual, we require for an
$N$ string amplitude $N-3$ of the operators of equation (6.2).  The
blocks $V(a,0)$ with $a=1,{15/16}$,  $V({1/2},0)$, $S$ and $P$
have ghosts number $3$, $2$, $-1$ and $1$ respectively.  The ghost number
requirement gives us the number $N_p$ of picture changing operators
$P$.  To be precise if we have a scattering of $N_1$ intercept $1$,
$N_{15/16}$ intercept ${15/16}$ and $N_{1/2}$ intercept
${1/2}$ strings, then $\varphi$ momentum conservation demands
that
$$
8N_1+7N_{15/16}+4N_{1/2}-2N_P = 14\eqno(6.3)
$$
while the ghost number count yields the relations
$$
3N_1+3N_{15/16}+2N_{1/2}-N_S+N_P-
(N_1+N_{15/16}+N_{1/2}-3)=8\eqno(6.4)
$$
These equations imply that
$$
\eqalign{&N_S=4\ N_1+{7\over2}N_{15/16}+2N_{1/2}-7\cr
&N_P=2\ N_1+{3\over2}N_{15/16}+N_{1/2}-2\cr}\eqno(6.5)
$$

We must now distribute all these factors of $P$ and some of the
factors of $S$ among the vertices so as to gain a non-zero result.
To be concrete let us consider the scattering of
$N_{15/16}=2n\geq 6$ intercept ${15/16}$ states, in which
case $N_S = 7(n-1)$ and $N_P=3n-2$.

We must assign these to the vertices of equations (5.21) and
(5.22).  One way to do this is to take $n-3$ of the vertices
$V({15/16},2)=(S^2 P)^2V({15/16},0)$ and $n+3$ of the
vertices $V({15/16},1)=S^{2}PV({15/16},0)$.  This leaves
over one $P$ factor and $n-1\ S$ factors and so leads to the
correlator
$$
\eqalign{\la0\vert&\prod\limits^{2n}_{i=4}\left\{\oint
dz_i\oint_{z_i}dv_i\
b(v_i)\right\}\prod\limits^{n+2}_{i=1}V({15/16},1)(z_i)\cr
&P\ V({15/16},1)(z_{n+3})\prod\limits^{2n}_{j=n4}\
V({15/16},2)(z_j)S^{n-1}\vert 0\ra\cr}\eqno(6.6)
$$
We have chosen to assign the final $P$ factor to the $n+3$
vertex, but any other vertex is just as good.  We could also use
two of the final screening charges on the $PV({15/16},1)$
vertex to yield the correlator
$$
\eqalign{&\la 0\vert\prod\limits^{2n}_{i=4}\left\{\oint
dz_i\oint\limits_{z_i}dv_i\
b(v_i)\right\}\prod\limits^{n+2}_{i=1}V({15/16},1)(z_i)\cr
&\prod\limits^{2n}_{j=n+3}V({15/16},2)(z_j)S^{n-
3}\vert 0\ra\cr}\eqno(6.7)
$$
Clearly there are many ways to write such a correlation using also
the vertices $V({15/16},m),m>2$.

For the scattering of $N_{1/2}=2n$ intercept ${1/2}$
states, $N_S=4n-7$ and $N_P=2(n-1)$.  In this case we can assign
all the factors of $S$ to the vertices, for example by the choice
of $n-2$ of the vertices $V({1/2},1)$, one of $\bar
V({1/2},0)$ and $n+1$ of the vertices $V({1/2},0)$ and then
place on one of these vertices the required  extra $P$.  The
correlator is then
$$
\eqalign{&\la0\vert\prod\limits^{2n}_{i=4}\left\{\int dz_i\oint dv_i
\,b(v_i)\right\}\prod\limits^{n+1}_{i=1}\ V({1/2},0)(z_i)\cr
&\prod\limits^{2n-1}_{j=n+2} V({1/2},1)(z_j)\ P\bar
V({1/2},0)(z_{2n})\vert 0\ra\cr}\eqno(6.8)$$

Finally for $N_1$ intercept 1 states, $N_S=4N_1-7$ and $N_P=2N_1-
2$, which we can assign as $N_1-2$ vertices $V(1,1)$, one of $V(1,0)$
and one of $\bar V(1,0)$, with one additional factor of $P$ to give
the correlator
$$
\eqalign{\la 0\vert\prod\limits^{N_1}_{i=4}\int
dz_i\oint\limits_{z_i}dv_i\ b(v_i)\prod\limits^{N_1-
2}_{j=1}V(1;1)(z_j)\cr &P\ V(1,0)(z_{N_1-1})\bar V(1,0)(z_{N_1})\
\vert 0\ra\cr}\eqno(6.9)$$

The many ways of constructing the above correlators lead to the
same results.  The correlator is independent of the place where the
picture changing operator is applied, since
$$
P(z_1)-P(z_2)=[Q,\varphi(z_1)-
\varphi(z_2)]=[Q,\int^{z_2}_{z_1}\partial\varphi],
$$
and this is a BRST trivial operator as it contains
$\partial\varphi$ and not $\varphi$.  Further, we can change on
which vertex the screening charges act by deforming the $w$-contours.
In fact there are different choices for the contours for the
residual screening charges and these lead to different results.  As
explained in section 4, in the context of the 4 intercept
${15/16}$ scattering, we require these different solutions
since only a particular combination of the contours gives a
crossing symmetric amplitude.

{\bf 7. Unitarity}

Unitarity of any theory requires that the $S$ matrix satisfy the
constraint $S^+S=1$ and that individual probabilities must be
positive.  The former condition implies the well known optical
theorem [25].  It is an obvious consequence of this theorem that
if the non-trivial part of the 2 particle to 2 particle scattering
amplitude were to vanish then the amplitude for the scattering of 2
particles to any number of particles would vanish.  This would
include the 2 particle to one particle scattering amplitude.

In references [21] and [22] a covariant approach to $W_3$
string scattering was given.  The authors adopted the usual rules
of the covariant scattering formalism for the bosonic string and
superstring, and applied them to the $W_3$ string.  They found that
many scattering amplitudes, particularly those involving intercept
${15/16}$ states, vanished.  They stressed the significance of
this fact and remarked that it contradicted the results of previous
work [17] and concluded that this former work, which gave all
$W_3$ scattering amplitudes, was incorrect.  In particular,
references [21] and [22] claimed that the scattering of 4
intercept ${15/16}$ string vanished, but that in agreement with
reference [17], the amplitude for 2 intercept ${15/16}$
strings to both intercept ${1/2}$ and 1 were non-vanishing.
This contradicts the optical theorem and so the set of
rules given in references [21] and [22] do not lead to a
unitary theory.

One might argue that, since the particles under discussion are unstable,
one could avoid this conclusion.  As one might expect, however, this
is not the case.
It can be shown [26], using only rather innocuous assumptions, that a 2
particle to 2 particle amplitude factorizes, near the energy of
the mass for a single intermediate state, into the product of the
two couplings of the 2 particles into the intermediate state
multiplied by
the intermediate state propagator.  The assumption of duality
[27] in string theory goes much further; it states that not only
can the entire amplitude be written as a sum over 3 particle
couplings multiplied by
the propagator of single particle intermediate states,
but that it can be
written in this way for either the $s$ or $u$ channel
exchanges.  Indeed it was this factorization procedure that
enabled the early pioneers of string theory to deduce all the
scattering amplitudes from the tachyon amplitudes.
As explained in this paper it is not sufficient just to  mimic the
covariant rules for the usual bosonic and superstrings,
but one should check whether a given set of string amplitudes factorizes
correctly.  The
covariant scattering formalism given in this paper  leads to a
non-vanishing amplitude for four intercept ${15/16}$ strings
which is in agreement with that found earlier[17].
In this reference it was indeed verified that the four ${15/16}$
intercept string amplitude does factorize correctly to give
three point functions that are
consistent with the fusion rules.  Thus the covariant scattering
formalism given in this paper is compatible with unitarity and with
the assumptions of S-matrix theory, as well as with
duality and factorization.
This is not the case for the covariant formalism of references [21] and
[22], which violate all of these principles.

As with any approach to string theory, other than gauge covariant
string field theory, unitarity must be used to fix the weights of
the individual amplitudes.

Although we have shown here that the approach advocated in
references [21,22] is not, in general, adequate to compute $W_3$
string scattering, nevertheless the amplitude for the scattering
of 4 tachyonic intercept ${1/2}$ states, does, as these authors
observed, agree with that found in reference [17].  In fact, all
of the non-vanishing scattering amplitudes found in references
[21] and [22] can readily be seen to agree with the general
formula for $W_3$ string scattering of reference [17].  In section 3
some further explicit evaluations of the general formula of
reference [17] were given to facilitate this comparison which
largely applies to the scattering of 4 tachyon strings.

The positivity of probabilities implies in particular that the norm
of physical states is positive.  It was shown in reference
[18] that the effective physical states for $c=51/2$ with
intercepts $1$, ${1/2}$ and ${15/16}$ had positive definite
norm.  Further, it was shown [18] that only for these three values of
the intercepts was this the case.  It was conjectured in reference
[18] that the cohomology of $Q$ consisted of these states,
copies of them and possible discrete states.  Considerable
additional evidence for this spectrum
conjecture was found in section 6.  It
follows from the conjecture and reference [18] that all physical
states of the $W_3$ string have positive norm and so satisfy a
no-ghost theorem.

It has been claimed in references [21] and [22] that one can
prove a no-ghost theorem by using the fact that the effective states
with $c=51/2$ at levels one and two have positive norm if the
intercept $a$ satisfies the bounds
$$
{15/16}\leq a\leq 1\quad {\rm or}\quad a\leq{1/2},$$
and that a standard result from ordinary string theory
is that the unitarity bounds derived from level 1 and level 2 states
are sufficient to ensure unitarity at all excited levels.

In fact in the early days of string theory positivity of the norm
up to quite high levels was investigated, but it required the
classic and beautiful work of reference [28] to
establish the no-ghost theorem, which is one of the cornerstones of
string theory.  Although the original arguments have been
simplified, as far as we are aware the shortest such arguments are
in essence the same as those given in the original paper.

In fact, at least as far as the evidence given in their paper is
concerned,  the authors of references [21] and [22] have
shown that demanding unitarity at levels 1 and 2 does
\underbar{not} imply unitarity at all levels since the former holds
for values of a other than $a=1,{15/16}$ and ${1/2}$. They do note,
however, that the bounds are saturated for the actual values of $a$.
In reference [18] it was shown
that an intercept not equal to
these values led to negative norm states at higher levels; it  might
perhaps be possible also to derive this result  by exploiting the fact
that the bounds are saturated for these values.

{\bf 8. Discussion}

In this paper we have given a general formalism for covariant
$W_3$ string scattering; any amplitude can be built from three
vertices $V(a,0)$, with $a = 1$, $15/16$ or $1/2$, together with
a screening charge $S$, a picture changing operator $P$, and the
usual $b$ ghost insertions.  The amplitudes are found to contain
Ising correlations and are in agreement with general results for
$W_3$ scattering found previously [17].  They are in disagreement,
however, with the results of references [21,22], which are shown to
violate the assumptions of S-matrix theory and the string assumption
of duality.

Starting from three states, one for each of the intercepts 1, 15/16
and 1/2, the screening and picture-changing operators are shown to
generate an infinite number of states in the cohomology of $Q$.
It is conjectured that these are all the elements of the cohomology
of $Q$, apart from discrete states and further states resulting from the
action of picture-changing operators.

It would be interesting to give a path-integral derivation of the
scattering results that we have obtained.  This would require a
knowledge of $W$-moduli.  Any such derivation would, however, have to
reproduce the results found in this paper, and this could provide a clue
to our understanding of $W$-moduli.  We observe that the number of
$W$-moduli is $2 N - 5$ for the scattering of $N$ strings, and this
number emerges from our results in the guise of $N_S - N_P + N_{1/2}$.

Although the scattering formalism given in references [21,22] was
not correct, these authors did propose an interesting field
redefinition.  Many of the calculations given in this paper are
considerably simplified when carried out in terms of these new
fields.

Although the string amplitudes we have found do contain the Ising model
correlation functions and we do use screening charges, these correlation
functions do not, at least at first sight, use the Feigin-Fuchs
representation.  It would be of interest further to investigate  this
phenomenon.

It would seem apparent that the pattern found for $W_3$ will also occur
for $W_N$.  The physical states will be effective space-time states with
intercepts given by weights in the corresponding unitary minimal models [18],
the scattering will contain the corresponding minimal model correlation
functions and the cohomology of $Q$ will be generated by acting with
screening charges and picture-changing operators on a set of states which
are in one to one correspondence with the weights in the corresponding
minimal models.

{\bf Acknowledgements}

We wish to thank N. Berkovits and G. Duke for discussions.
M. Freeman is grateful to SERC for financial support.

{\bf Appendix A}

In this appendix we summarize some of the known technical results
[29] used to compute the scattering amplitudes of the $W_3$
string which involve the hypergeometric function
$F(a,b,c;z)=F(b,a,c;z).$  This function is known to possess the
integral representation
$$
F(a,b;,c,z)={\Gamma(c)\over\Gamma(b)\Gamma(c-b)}\int^1_0 dt\,t^{b-
1}(1-t)^{c-b-1}(1-tz)^{-a}\eqno(A1)
$$
By the change of variable $w={1/t}$ and redefinition
of a,b and c we find that
$$
\eqalign{&\int^\infty_1 w^c(w-1)^b(w-z)^a dw\cr &={\Gamma(-1-b-c-
a)\Gamma(b+1)\over \Gamma(-c-a)} F(-a,-b-c-a-1;c-
a,z)\cr}\eqno(A2)
$$
Similarly, one finds that
$$
\eqalign{&\int^z_0 dw w^a(1-w)^b(z-
w)^c=z^{1+a+c}{\Gamma(a+1)\Gamma(c+1)\over\Gamma(a+c+2)}\cr &F(-
b,a+1,a+c+2;z)\cr}\eqno(A3)
$$
For the four intercept ${15/16}$ strings one finds certain
hypergeometric functions that can be given in terms of elementary
functions by the following identities
$$
\eqalign{\cos a\theta &=F({a\over2},-
{a\over2},{1\over2};\sin^2\theta)\cr &=\cos\theta F({a+1\over2},{1-
a\over2},{1\over2};\sin^2\theta)\cr \sin a\theta&=a\sin\theta
F({a+1\over2},{1-a\over2},3/2;\sin^2\theta)\cr &=a\sin\theta \cos\theta
F(1+{a\over2},1-{a\over2},3/2;\sin^2\theta)\cr}.\eqno(A4)$$

{{\bf References}}
\parskip 0pt
\item{[1]} A. B. Zamolodchikov, Theor. Math. Phys 65 (1989) 1205.
\item{[2]} V.A. Fateev and S. K. Lukyanov, Int. J. Mod. Phys. A3 (1988) 507.
\item{[3]} J. Thierry-Mieg, Phys. Lett. B197 (1987) 368.
\item{[4]} A. Bilal and J. L .Gervais, Nucl. Phys. B326 (1989) 222.
\item{[5]} P. Howe and P .West, unpublished
\item{[6]} V .A .Fateev and A. B. Zamolodchikov,
Nucl. Phys. B280 [FS18] (1987) 644.
\item{[7]} L. J. Romans , Nucl. phys {\bf B352 } (1991) 829.
\item{[8]} S. Das, A .Dhar and S .Kalyana Rama, Mod. Phys. Lett. 268B (1991);
Mod. Phys. Lett. 269B (1991) 167; Int.J. Mod. Phys. A7 (1992) 2295.
\item{[9]} C. N. Pope, L .J. Romans, K. S .Stelle, Phys. Lett. 268B (1991) 167;
Phys. Lett. 269B (1991) 287.
\item{[10]} H. Lu, C. N. Pope, S. Schrans and K. W. Xu,
Texas A \& M preprint CTP TAMU-5/92.
\item{[11]} E. del Giudice and P. di Vecchia, Nuovo Cimento 5A (1971) 90.
\item{[12]} M. Kato and K. Ogawa, Nucl. Phys. B212 (1983) 443.
M. D. Freeman and D. I. Olive, Phys. Lett. B 175 (1986) 151.
\item{[13]} H. Lu, B. E. W. Nilsson, C. N. Pope, K. S. Stelle and P. West,
``The low level spectrum of the $W_3$ string,''
\item{[14]} S. Kalyana Rama ,Mod. Phys. Lett {\bf A}6 (1991)3531.
\item{[15]} B. H. Lian and G. J. Zuckerman, Phys. Lett. B 254 (1991);
Phys. Lett. B 266 (1991) 21.
A. M. Polyakov, Mod. Phys. Lett. A6 (1991) 635.
E. Witten, Nucl. Phys. B 373 (1992) 187.
N. Ohta, ``Discrete states in two-dimensional quantum gravity''
and references therein.
\item{[16]} C. N.  Pope, E. Sezgin, K. S. Stelle and X. J. Wang,
``Discrete states in the $W_3$ string,'' CTP-TAMU-64/92, Imperial/TP/91-92/40.
\item{[17]} M. Freeman and P. West;  "$W_3$ string scattering", KCL-TH-92-4,
NI-92007,Phys. Lett. {\bf B} to be published.
\item{[18]} P. West, ``On the spectrum, no-ghost theorem and modular
invariance of $W_3$ strings,'' KCL-th-92-7, to be published in
Int. J. Mod. Phys.
\item{[19]} For a review see D. Lust and S. Theisen, ``Lectures on string
theory,'' Springer-Verlag 1989.
\item{[20]} F. L. Feigin and D. B. Fuchs, Moscow preprint 1983.
Vl. S. Dotsenko and V. A. Fateev, Nucl. Phys. B240 (1984) 312.
\item{[21]} H. Lu, C. N. Pope, S. Schrans and X-J. Wang, ``The
interacting $W_3$ string,'' CTP-TAMU-86/92.
\item{[22]} H. Lu, C. N. Pope, S. Schrans and X-J. Wang, ``On the spectrum
and scattering of $W_3$ strings,'' CTP-TAMU-4/93.
\item{[23]} E. Witten and  B. Zwieback, Nucl. Phys. {\bf B377} (1992) 644.
\item{[24]} For a review see C. Itzykson and M. Drouffe, ``Statistical
Field Theory,'' Cambridge University Press.
\item{[25]} See, for example, C. Itzykson and J-B. Zuber,
``Quantum Field Theory,'' McGraw Hill.
\item{[26]} R. Eden, P. Landshoff, D. Olive and J. Polkinghorne,
``The Analytic S-matrix,'' Cambridge University Press.
\item{[27]} R. Dolan, D. Horn and C. Schmid, Phys. Rev. Lett. 19 (1967)
402; Phys. Rev. Lett. 166 (1968) 1768.
\item{[28]} R .C. Brower and P. Goddard, Nucl. Phys.B40 (1972) 437;
R .C .Brower, Phys. Rev. D6 (1972) 1655; P .Goddard and C. Thorn,
Phys. Lett. 40B (1972) 235.
\item{[29]} ``Bateman Manuscript Project,'' edited A. Erdelyi,
McGraw Hill.
%
%
%
%
%
%
%
%
%
%
%
%
%

\end